\documentclass[twocolumn]{aastex631}
\usepackage{amsmath}
\submitjournal{ApJ}

\begin{document}
    
\title{ UV Spectral Slope and Nebular Dust Attenuation in Dwarf Galaxies at $1.4<z<2.6$}

\correspondingauthor{Anahita Alavi}
\email{anahita@ipac.caltech.edu}

\author[0000-0002-0786-7307]{Anahita Alavi}
\affiliation{IPAC, Mail Code 314-6, California Institute of Technology, 1200 E. California Blvd., Pasadena CA, 91125, USA}

\author[0000-0002-4935-9511]{Brian Siana}
\affiliation{Department of Physics \& Astronomy, University of California, Riverside, CA 92521, USA}

\author[0000-0002-7064-5424]{Harry I. Teplitz}
\affil{IPAC, Mail Code 314-6, California Institute of Technology, 1200 E. California Blvd., Pasadena CA, 91125, USA}

\author[0000-0002-7732-9205]{Timothy Gburek}
\affiliation{Department of Physics \& Astronomy, University of California, Riverside, CA 92521, USA}

\author[0000-0001-6482-3020]{James Colbert}
\affiliation{IPAC, Mail Code 314-6, California Institute of Technology, 1200 E. California Blvd., Pasadena CA, 91125, USA}

\author[0000-0001-7166-6035]{Vihang Mehta}
\affiliation{IPAC, Mail Code 314-6, California Institute of Technology, 1200 E. California Blvd., Pasadena CA, 91125, USA}

\author[0000-0003-2047-1689]{Najmeh Emami}
\affiliation{Minnesota Institute for Astrophysics, University of Minnesota, Minneapolis, MN, 55455, USA}

\author[0000-0003-3559-5270]{William R. Freeman}
\affiliation{Department of Physics \& Astronomy, University of California, Riverside, CA 92521, USA}

\author[0000-0001-5492-1049]{Johan Richard}
\affiliation{Univ Lyon, Univ Lyon1, Ens de Lyon, CNRS, Centre de Recherche Astrophysique de Lyon UMR5574, F-69230, Saint-Genis-Laval, France}

\author[0000-0001-6505-0293]{Keunho Kim}
\affil{IPAC, Mail Code 314-6, California Institute of Technology, 1200 E. California Blvd., Pasadena CA, 91125, USA}



\begin{abstract}

We analyze nebular dust attenuation and its correlation with stellar mass ($M_{*}$) and UV spectral slope ($\beta$) in 33 lensed, low-mass star-forming galaxies at $1.4\leq z \leq 2.6$, using Keck/MOSFIRE rest-frame optical spectroscopy. Located behind three massive lensing galaxy clusters Abell 1689, MACS J1149.5+2223, and MACS J0717.5+3745, galaxies in our sample have a median  stellar mass of $\log(M_{*}/M_{\odot})=8.3$ and an intrinsic UV absolute magnitude range of $-20.9<M_{UV}<-13$. We measure nebular dust attenuation via Balmer optical depth ($\tau_{B}$) defined as the H$\alpha$/H$\beta$ ratio. We also derive physical properties from Hubble Space Telescope multi-wavelength photometry and construct composite spectra using median stacking in bins of $M_{*}$ and $\beta$.
We find that the $\tau_{B}-\beta$ relation for the dwarf galaxies in this study is best represented by SMC dust curve. 
This is consistent with previous studies of low-metallicity galaxies at similar redshifts, which show a steep attenuation curve similar to the SMC curve, in contrast to high-metallicity and more massive galaxies that exhibit a much shallower dust attenuation curve.
We also investigate the relationship between nebular dust attenuation and stellar mass, $E(B-V)_{nebular}-M_{*}$, down to $\log(M_{*}/M_{\odot})\sim 7$. We demonstrate that this relation does not notably evolve with redshift and is consistent with what has been observed for local SDSS galaxies at similar low stellar masses.

\end{abstract}



\section{Introduction} 
\label{sec:intro}

Dust grains in galaxies are not just minor components; they are crucial in shaping our understanding of galaxies' physical properties, formation, and evolution. Dust absorbs and scatters light over a wide range of ultraviolet (UV) and optical wavelengths, then re-emits it in infrared (IR). This process affects the observed spectrum of galaxies, thereby influencing our interpretation of their observable characteristics.

In order to recover the intrinsic physical properties of galaxies, such as star formation rate (SFR), age, and metallicity, one must correct for dust attenuation. There are various techniques that use light at different wavelengths to measure dust attenuation in galaxies. 
Far-Infrared (FIR) luminosity of galaxies is dominated by dust re-radiating the absorbed stellar energy from UV-optical wavelengths. Therefore, as shown in many studies \citep[e.g., ][]{cal00,xu1996}, for star-forming galaxies, the FIR luminosity to UV ratio 
(i.e., IRX) is an estimate of the global dust optical depth and attenuation at UV wavelength (i.e., $A_{UV}$).  \citet{meurer1999} showed that there is a correlation between the IRX ratio and UV spectral slope $\beta$ for local starburst galaxies. 
\textbf{Later studies showed that this relation and the position of normal star-forming galaxies in this parameter space is correlated with several factors such as their underlying dust attenuation curve \citep{sia09, salim2020}, metallicity and environment \citep{Shivaei2020a}, recent star formation \citep{Casey2014}, and mixed of different parameters \citep{Buat2012}. In addition, several studies using simulations and models have also investigated the origin and scatter observed in the IRX-$\beta$ relation \citep{Popping2017,Safarzadeh2017,Narayan2018,Schulz2020}}.
\textbf{Considering this strong correlation between UV spectral slope and dust, $\beta$ has been the most popular proxy for measuring dust attenuation for high redshift galaxies} where observing the rest frame FIR luminosity is not feasible \citep [e.g.,] [] {cal94, reddy12a,bou12b, red18, cullen2023,cullen2024}.
Besides utilizing continuum light, spectroscopy offers a reliable means of measuring dust. 

In rest-optical wavelengths, the ratio of nebular emission lines, primarily H$\alpha$/H$\beta$, known as the Balmer decrement, has been extensively used to estimate dust attenuation in local galaxies \citep{kennicutt1992, groves2012, Qin2019}. \textbf{This method has also been applied to galaxies at cosmic noon \citep{dominguez2013, kashino2013, Nelson2016,Theios2019, Shivaei2020b, shapley2022, Mathru2023}} and, more recently, to very high-redshift galaxies at $z=3-7$ using James Webb Space Telescope (JWST) spectroscopy \citep{shapley2023, sandles2023}.


Various studies have investigated the dependence of dust attenuation, measured using different techniques listed above, on multiple galaxy properties. \citet{maheson2024} demonstrated that for a large sample of local SDSS galaxies, dust attenuation measured using the Balmer decrement primarily depends on stellar mass, followed by metallicity and velocity dispersion. The strong correlation between dust attenuation and stellar mass for star-forming galaxies has been observed in numerous other studies as well. One interesting observation is that this strong correlation does not evolve with redshift. This finding holds true regardless of the method used to measure dust attenuation. 
\citet{whitaker2017} parameterized dust attenuation as the fraction of obscured star formation rate (SFR), defined as SFR$_{IR}$/SFR$_{UV+IR}$. They found a strong dependence of the fraction of obscured SFR on stellar mass, with remarkably little evolution in this fraction with redshift out to $z=2.5$. \textbf{A recent study by \citet{Shivaei2024} from JWST MIRI observations also shows the same results.}  Using another method of measuring dust attenuation, the IRX ratio, several studies \citep{heinis2014, bouwens2016, Mclure2018, alvarez2019} demonstrated that the IRX-$M_{*}$ relation does not evolve with redshift up to $z=3-4$. Dust attenuation measured at UV wavelengths ($1600$ \AA, A$_{1600}$ ) \citep{pannella2015} and similarly measured via SED fitting \citep{martis2016,shapley2022} also shows no evolution with redshifts up to $z=3-4$. 
Finally, the lack of evolution in the relationship between nebular dust attenuation (i.e., Balmer decrement) and $M^{*}$ has also been reported in previous works such as \citet{dominguez2013, kashino2013, price2014}, as well as in a more recent study by \citet{shapley2022} up to $z=2$. Another work by \citet{shapley2023} using JWST/NIRSpec spectroscopy also demonstrates that the lack of evolution in nebular attenuation at fixed stellar mass holds up to much higher redshift, at $z=6.5$. It is important to note that almost all of these studies at cosmic noon have drawn their conclusions for star-forming galaxies with $M*>10^{9} \ M_{\odot}$.
 
 In this paper, we build a sample of 33 low-mass star-forming galaxies with $M*<10^{9} \ M_{\odot}$ at $1.43<z<2.66$ with high quality Near-IR spectra from Keck/MOSFIRE \citep{mclean2012} as well as deep UV imaging from HST. These galaxies are gravitationally lensed and therefore we can probe very low masses. In Section \ref{sec:data}, we describe our dataset and introduce our sample selection. Utilizing the rich dataset, we then perform SED fitting for galaxies in our sample and measure their physical properties including their UV spectral slope in Section \ref{sec:sed-fit} and \ref{sec:uv-slope}. We analyze dust attenuation based on H$\alpha$/H$\beta$ Balmer line ratio using the rest-frame optical spectra from MOSFIRE for both individual galaxies and stack in Section \ref{sec:nebular_dust} and Section \ref{sec:stack}. The results are presented in Section \ref{sec:results}.

In this paper, all magnitudes are in AB system \citep{oke83}. We adopt 
$\Omega_{M}=0.3$, $\Omega_{\Lambda}=0.7$ and $H_{0}=70 \ \text{km} \ \text{s}^{-1} \ \text{Mpc}^{-1}$.

\section{Data and Sample Selection}
\subsection{photometric data and catalogs}
The main sample in this paper originates from spectroscopic follow-up of an HST photometric survey conducted across three lensing galaxy clusters: Abell 1689 (A1689), MACS J1149.5+2223 (M1149), and MACS J0717.5+3745 (M0717). A comprehensive description of the HST imaging survey, photometric data, and catalogs can be found in \citet{ala14} (hereafter A14) and \citet{ala16} (hereafter A16). In the following sections, we provide a brief overview of the photometric data and catalogs, followed by an explanation of how our spectroscopic samples were constructed.

For the two clusters M1149 and M0717, which are part of the Hubble Frontier Fields program \citep{lot16}, we utilize publicly available deep data in 9 {\it HST} broad-band filters, including F275W, F336W, F435W, F606W, F814W, F105W, F125W, F140W, and F160W. The UV data are sourced from {\it HST} program ID 13389 (PI: Siana), while the optical and NIR data are obtained from {\it HST} program ID 13495 (PI: Lotz). For A1689, we utilize publicly accessible photometric data in 7 {\it HST} bands, specifically  F275W, F336W, F475W, F625W, F775W, F814W, and F850LP. The UV data for A1689 are associated with program IDs 12201 and 12931 (PI: B. Siana).

With the available photometric data, we utilized the \texttt{EAZY} code \citep{bram08} as detailed in A16 to estimate photometric redshifts, enabling us to construct a sample of star-forming galaxies within the redshift range $1<z<3$. Leveraging the magnification from strong gravitational lensing by the foreground galaxy clusters, as discussed in A16, we were able to probe extremely faint luminosities, down to an intrinsic UV magnitude of $M_{UV}=-13$. Furthermore, the deep near-UV bands in these fields enabled us to identify the Lyman breaks and accurately measure the photometric redshifts.
 As described in A16, to correct for the lensing magnification and estimate the intrinsic physical properties of galaxies, we used the lens models from \citet{lim07,jauzac2016,lim16} for A1689, M1149 and M0717, respectively.

\label{sec:data}	
\subsection{Spectroscopic data }
Following the {\it HST} photometric observations detailed in A16, we conducted a spectroscopic survey between 2014 and 2017 to acquire near-infrared (rest-frame optical) spectroscopy for a sample of lensed, faint star-forming galaxies within the redshift range of $1<z<3$. We targeted highly magnified galaxies behind A1689, M0717, and M1149. This was accomplished using the Multi-Object Spectrometer For Infrared Exploration \citep[MOSFIRE, ][]{mclean2012} mounted on the 10 m Keck I telescope. 

\textbf{The selection of the spectroscopic targets was based on specific criteria. First, the targets were chosen to fall within the photometric redshift range of about $1 < z < 3$, allowing for the detection of rest-frame optical nebular emission lines.}
\textbf{A full description of the photometric redshift measurements is provided in \citet{ala16} (also see Section \ref{sec:data}), where the redshifts are primarily estimated using the Lyman break. Although, our UV filters where the Lyman break falls is very deep, but this may still introduce some bias in our selection, as galaxies with very weak Lyman breaks, such as very dusty or evolved systems, could be underrepresented.} The targets were then required to be sufficiently bright, with an observed optical magnitude (m$_{B}$) below 26.5, ensuring a high signal-to-noise ratio for line detection. Moreover, we prioritized targets with high lensing magnification to facilitate the investigation of intrinsically fainter sources.

We explicitly targeted the strong nebular emission lines of [OII] $\lambda\lambda$ 3726, 3729, H$\beta$, [OIII] $\lambda\lambda$ 4959, 5007, H$\alpha$, and [N II] $\lambda\lambda$ 6548, 6583. To accomplish this goal, we conducted observations of galaxies in the lowest redshift range using the Y-, J-, and H-band filters, while the J-, H-, and K-band filters were used for the two highest redshift ranges. Depending on the band, redshift and observation conditions (e.g., seeing), the individual exposure times varied between 48-120 minutes. 
A detailed description of the spectroscopic data reduction, encompassing wavelength and flux calibration, can be found in \citet{gburek23} and \citet{emami20}. Here, we provide a brief overview of the key aspects. The MOSFIRE spectroscopic observations were taken in an ABBA dither pattern with a $2\farcs5$ dither spacing. The 2D spectroscopic data reduction was performed using the MOSFIRE Data Reduction Pipeline (DRP \footnote{https://keck-datareductionpipelines.github.io/MosfireDRP/}). This pipeline produces a composite spectrum from multiple spectra captured at each dither position and includes flat-fielding, wavelength calibration, background subtraction, and rectification. Subsequently, the 1D spectral extraction was carried out employing the BMEP code developed by \citet{freeman2019}. For the 1D flux calibration, we adopted a procedure similar to that outlined by \citet{kriek2015}. First, we conducted an initial flux calibration utilizing a B9 V to A2 V standard star observed at the same airmass as our target. Subsequently, we performed an absolute flux calibration using a slit star observed within the same slit mask as our science targets.

\subsection{Line Flux Measurement}
\label{line_flux}
A detailed description of the spectroscopic line measurements and fitting procedures can be found in \citet{gburek23}. Here, we provide a summary of the techniques they used. The emission line spectra are fitted using the Markov Chain Monte Carlo Ensemble sampler, emcee \citep{foreman2013}. In this approach, a general model is employed, consisting of a line fit to the continuum and a single Gaussian profile for emission lines such as H$\alpha$, H$\beta$, [OIII] $\lambda\lambda$ 4959, 5007. To mitigate the impact of sky lines on the data, we exclude pixels with errors exceeding $3\times$ the median error value over the range of the fit prior to line fitting. It's important to note that the spectrum in each band (Y, J, H, and K) is fit independently. Ultimately, the final redshift estimate is determined as the weighted average of the redshifts fit in each band. Additionally, the flux of an emission line is calculated using the formula $f= \sqrt{2\pi}A\sigma$, where $A$ represents the emission line amplitude and $\sigma$ denotes the line width.

In slit spectroscopy, line fluxes must be corrected for slit losses as the slit may not fully encompass the object. Our emission line fluxes are corrected for slit losses using the method outlined in detail in \citet{emami20}.

Finally, our measured Balmer emission line fluxes are likely affected by Balmer absorption lines from the atmosphere of stars, mainly of A-type. These absorption lines are present in the real stellar continuum of each spectrum. We estimate the Balmer absorption lines from the best SED fits described in Section \ref{sec:sed-fit}. The mean stellar absorption corrections in our sample for H$\alpha$ and H$\beta$ are $1.3\%$ and $8.8\%$, respectively. These values are consistent to what has been reported in \citet{reddy2015} and \citet{Kashino2019} for the Balmer absorption corrections of their samples. 



\subsection{Sample selection}
Among our spectroscopic targets, there are 32, 13, and 6 sources within the A1689, M0717, and M1149 clusters, respectively, where spectra cover both H$\alpha$ and H$\beta$ wavelengths. This coverage is crucial for the science of this paper, which focuses on measuring nebular dust attenuation. Following this, we select sources with 3$\sigma$ detected H$\alpha$ emission line. Consequently, the counts of sources within the A1689, M0717, and M1149 clusters drop to 28, 11, and 6, respectively. Furthermore, to ensure reliable emission line flux measurements, especially for H$\beta$, we exclude sources under the following conditions:

1- Significant contamination from sky lines at the wavelengths of observed H$\alpha$ or H$\beta$ emission lines.

2- Sources that are excessively large and extended, with slit loss corrections exceeding $70\%$. We note that we are not excluding galaxies with large effective radii, but instead this is generally removing large arcs.

Due to the effects of strong gravitational lensing, some galaxies in our sample exhibit multiple images. To avoid double counting, we exclude these multiple images from our analysis. 
Ultimately, our final sample comprises 21, 6, and 6 sources within the A1689, M0717, and M1149 clusters, respectively.

\begin{figure*}[ht]
    \includegraphics[width=\textwidth]{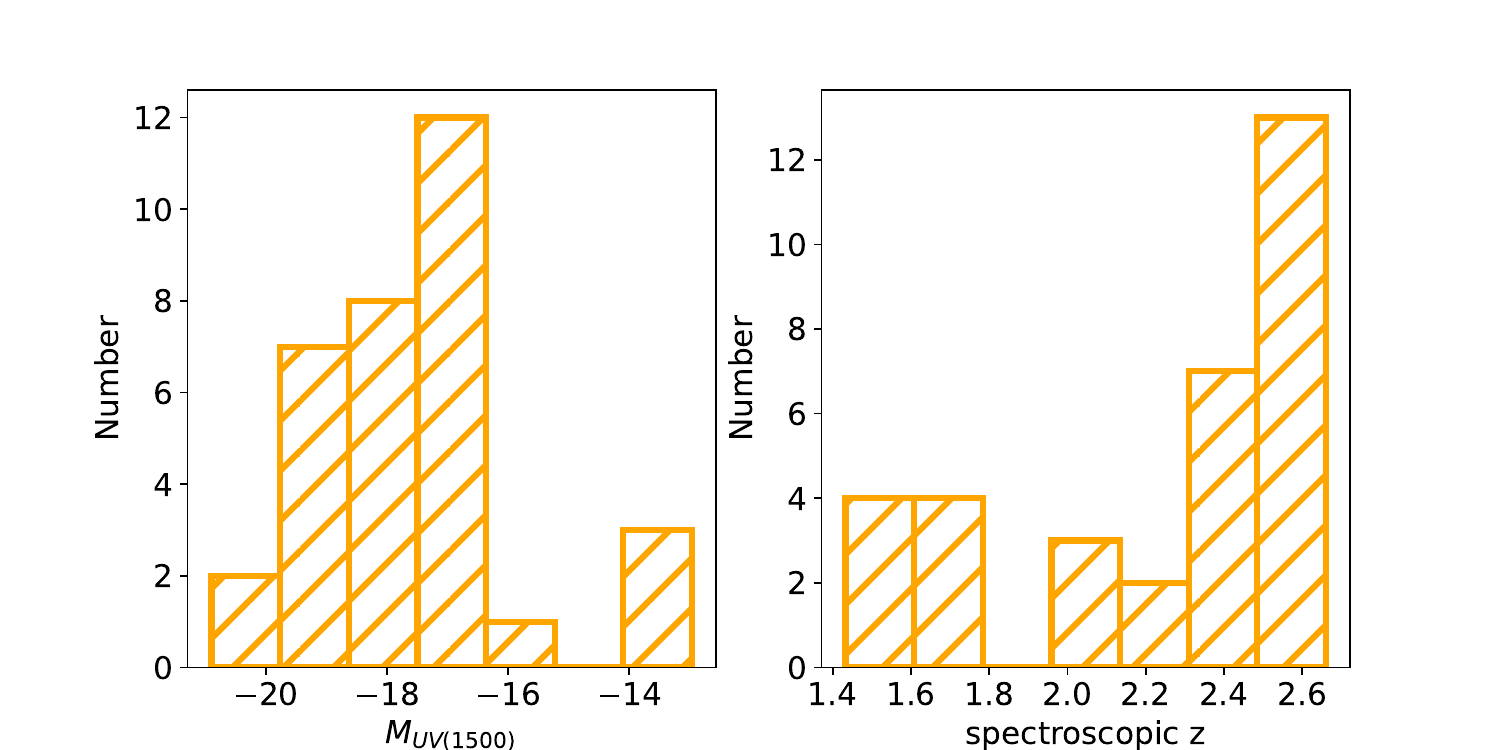}
    \caption{Left: Distribution of the rest-frame UV absolute magnitudes at 1500 \AA\ for our final sample. These $M_{UV}$ values are corrected for the lensing magnification. Right: Distribution of spectroscopic redshifts of our sample. }
\label{fig:Muv_z_hist}
\end{figure*}

In the right panel of Figure \ref{fig:Muv_z_hist}, we illustrate the distribution of spectroscopic redshifts of sources in our sample. As seen, our sample spans a redshift range of $1.43<z<2.66$. Following the methodologies outlined in A16, we utilize the {\it HST} broad-band imaging in F336W, F435W, and F606W bands for the HFFs, and the F336W, F475W, and F625W bands for the A1689 samples to measure the UV absolute magnitude at rest-frame 1500 \AA\ at redshifts $1.4<z<1.6$, $1.6<z<2.2$, and $2.2<z<2.6$, respectively. The distribution of the rest-frame UV absolute magnitudes at 1500 \AA\ is depicted in the left panel of Figure \ref{fig:Muv_z_hist}. We note that these magnitudes are corrected for the lensing magnification, and are therefore intrinsic.

\section{SED Fitting}
\label{sec:sed-fit}
To measure physical properties of galaxies in our sample, such as stellar mass and UV spectral slopes, we perform SED fitting to multi-band photometry using the code Fitting and Assessment of Synthetic Templates (\texttt{FAST}) \citep{kri09}. \texttt{FAST} utilizes the synthetic stellar population models from \citet{bru03} (hereafter BC03) and determines the best-fit SED through a $\chi^2$ fitting process. For the input parameters to \texttt{FAST}, we assume a Chabrier initial mass function (IMF) \citep{cha03} and consider the metallicity as a free parameter between [$0.2\ Z_{\sun}$ , $0.4\ Z_{\sun}$] values. 
\textbf{There are two reasons which justify these metallicity assumptions. First, as described in Section \ref{subsec:balmer_depth_UV}, we have direct metallicity measurements for a sub-sample of galaxies in this study using the [OIII] $\lambda 4363$ auroral line \citep{gburek23}. These measurements yield an average metallicity of $0.15 Z_\odot$, providing an empirical basis for the assumed metallicity range.
Second, considering the well-established mass-metallicity relations from the literature, both recent and earlier works (e.g., Li et al. 2023; Erb et al. 2006; Belli et al. 2013), galaxies with stellar masses in the range of $7<\log(M*/M_{\odot})<9$ are typically found to have gas-phase metallicities around $0.2-0.4 \ Z_\odot$. Assuming that the stellar and gas-phase metallicities are similar, this supports our adopted range as being representative of the expected metallicities for our sample. Lastly, we note that the primary physical parameter derived from the SED fitting in this work is the stellar mass. It is well established that the assumed metallicity has only a minor impact on stellar mass estimates. To verify this, we re-ran our SED fitting with a broader metallicity range ($0.2-1.0 \ Z_\odot$), and found that the resulting stellar masses remained in very close agreement with our original estimates.}

As justified in \citet{reddy12a} for high-redshift galaxies at $z>2$, we adopt an exponentially increasing star formation history (i.e., SFH $\propto e^{t/\tau}$) with $8<\log(\tau)<11$. The age timescale can also vary between $7<\log(t)[\text{yr}]<10$. We assume a Calzetti dust attenuation curve with $0<A_{v}<3$. The redshifts are fixed to the spectroscopic redshifts of the sources for the SED fitting process.

\begin{figure}[t]
\label{fig:mass_hist}
    \includegraphics[width=\columnwidth]{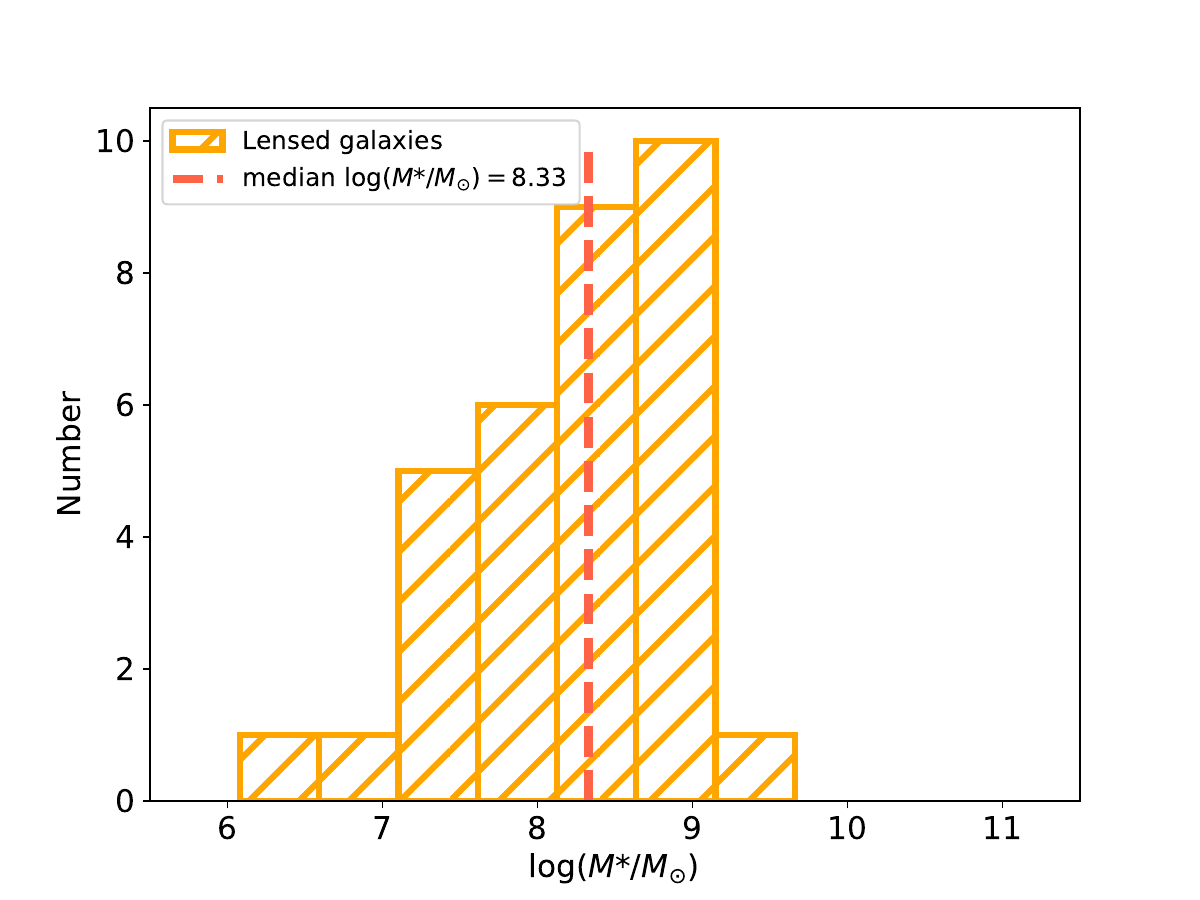}
    \caption{Histogram of stellar mass $M^{*}$ for the 33 galaxies in our final sample. These mass values are from SED fitting and are corrected for magnification. }
\end{figure}

Several studies at high redshifts \citep[e.g.,][]{de-barros2014} have suggested that emission lines with high equivalent widths, particularly in lower-mass galaxies like those in our sample, can significantly impact the determination of physical parameters from SED fitting. Therefore, before conducting SED fitting, we subtracted the contribution of slit-loss-corrected nebular emission lines from the photometry. Figure \ref{fig:mass_hist} illustrates the stellar mass (magnification corrected) distribution of our sample obtained from the SED fitting. The galaxies in our sample span a stellar mass range of $6.1<\log(M^{}/M_{\odot})<9.7$, with a median value of $8.3$. 


\section{UV Spectral Slope}
\label{sec:uv-slope}
The rest-frame UV continuum spectra of galaxies can be parameterized by a power-law function as $f_{\lambda} \propto \lambda^{\beta}$,
where $f_{\lambda}$ is the flux density per wavelength and $\beta$ is the UV continuum slope \citep{cal94}. As defined by \citet{cal94}, we adopt the wavelength range of $\lambda=1250 - 2600$ \AA\ for our UV slope measurements. This parametrization was originally presented for the UV continuum spectroscopy of galaxies. However, the $\beta$ values of high-redshift galaxies are often measured from broad-band photometry due to the lack of rest-frame UV spectroscopic data. In this section, we describe two methodologies and explore their relative merit for measuring the UV spectral slopes of our lensed galaxies at $1.4<z<2.6$.

\subsection{UV slope: Power Law Fitting}
In this technique, which has been utilized in numerous previous studies \citep[e.g.,][]{bou09, fin12} and more recently for JWST data \citep[e.g.,][]{cullen2023, Morales2023}, we use the redshift of each galaxy to select the broad-band filters within the rest-frame wavelength range of $\lambda=1250 - 2600$ \AA\ \citep[Calzetti wavelength range; ][]{cal94}. A broad-band filter is considered if the mean wavelength 
\footnote{The mean wavelength is defined as 
$\lambda_{\text{mean}} = \frac{\int \lambda T_{\lambda} \, d\lambda}{\int T_{\lambda} \, d\lambda}$.}
of the corresponding filter falls within this wavelength range.  Furthermore, to avoid contamination from the Ly$\alpha$ emission line, we exclude any filter whose transmission at the redshifted Ly$\alpha$ wavelength exceeds $50\%$ of its maximum transmission. Consequently, at each redshift, only a specific group of filters satisfies all these criteria.

\textbf{We assume a power-law function, $f_{\lambda} = \lambda^{\beta}$, for the rest-frame UV continuum spectrum (e.g., SED) of each galaxy. We then convolve this power-law function with the filter transmission curves to compute synthetic fluxes for each selected band within the Calzetti wavelength range (see above). Next, we calculate synthetic colors using these synthetic fluxes of each pair of photometric bands within the Calzetti wavelength range. By minimizing $\chi^{2}(\beta)$ between the synthetic and observed colors of each source, we recover the best-fit $\beta$. This procedure yields the same result as a direct power-law fit to $f_\lambda$. In the direct approach, one converts the broadband magnitudes to flux densities ($f_{\nu}$), assumes the central (or effective) wavelength of each filter to calculate $f_{\lambda}$, and then fits a $\lambda^{\beta}$ function to the points. However, our approach has the advantage of naturally accounting for the full filter transmission profiles.}

 \textbf{An advantage of power-law technique is that the uncertainty estimates are data-driven, derived directly from the photometric uncertainties.
However, due to the limited number of passbands, some of our $\beta$ values are derived from only single-color measurements, leading to larger uncertainties. Henceforth, we will refer to these measurements as $\beta_{\mathrm{power}}$.}


\subsection{UV slope: SED Fitting}
\label{beta-sed}
Another approach for measuring the $\beta$ slope involves making full use of the available photometry \citep{fin12, rogers2013}. This entails an observed wavelength coverage of 0.27-1.54 $\mu$m for M0717 and M1149 and 0.24-0.91 $\mu$m for A1689. In this technique, we determine $\beta$ by fitting the best SED of each galaxy derived in Section \ref{sec:sed-fit}.

    
To avoid the effect of absorption and emission features, \citet{cal94} proposed using 10 spectral windows (i.e., Calzetti windows) between 1250-2600 \AA\ for $\beta$ measurements. Utilizing the best-fit SED of each object, we compute the mean wavelength and flux within these windows. Subsequently, we determine $\beta$ through a linear fit to log$(f_{\lambda})$ versus log$(\lambda)$. We estimate the uncertainties of $\beta$ by conducting a series of Monte Carlo simulations. In these simulations, the observed fluxes for each source are randomly perturbed based on their uncertainties. We then rerun \texttt{FAST}, refit the $\beta$ values, and calculate the $68\%$ confidence interval. Throughout the remainder of the paper, we will refer to these measurements as $\beta_{\mathrm{SED}}$. 

\textbf{We note that our $\beta_{\mathrm{SED}}$ measurements are observed UV slopes and are independent of the assumed dust attenuation curve in the SED fitting. As demonstrated in other studies \citep{Lafaro2017, Salmon2016}, this is due to the fact that the best-fitting SED always provides a close match to the UV colors as long as the assumed dust law does not have extreme features.}
\textbf{An advantage of the SED fitting method compared to the power-law fitting is that $\beta_{\mathrm{SED}}$ is constrained using the full wavelength range between 1250–2600 \AA, whereas $\beta_{\mathrm{power}}$ relies on only a few discrete colors within this range. As a result, $\beta_{\mathrm{SED}}$ is less affected by absorption and emission features in the broad-band photometry \citep{cal94,rogers2013}. However, a disadvantage of the SED-fitting method is that the uncertainties can be optimistic. In addition, the $\beta_{\mathrm{SED}}$ measurements are constrained by the allowable values from the stellar population models used for the SED fitting.}

\subsection{Comparing the Methods}

\begin{figure}[ht]
\includegraphics[trim= 0cm 4cm 0cm 0cm,,width=\columnwidth]{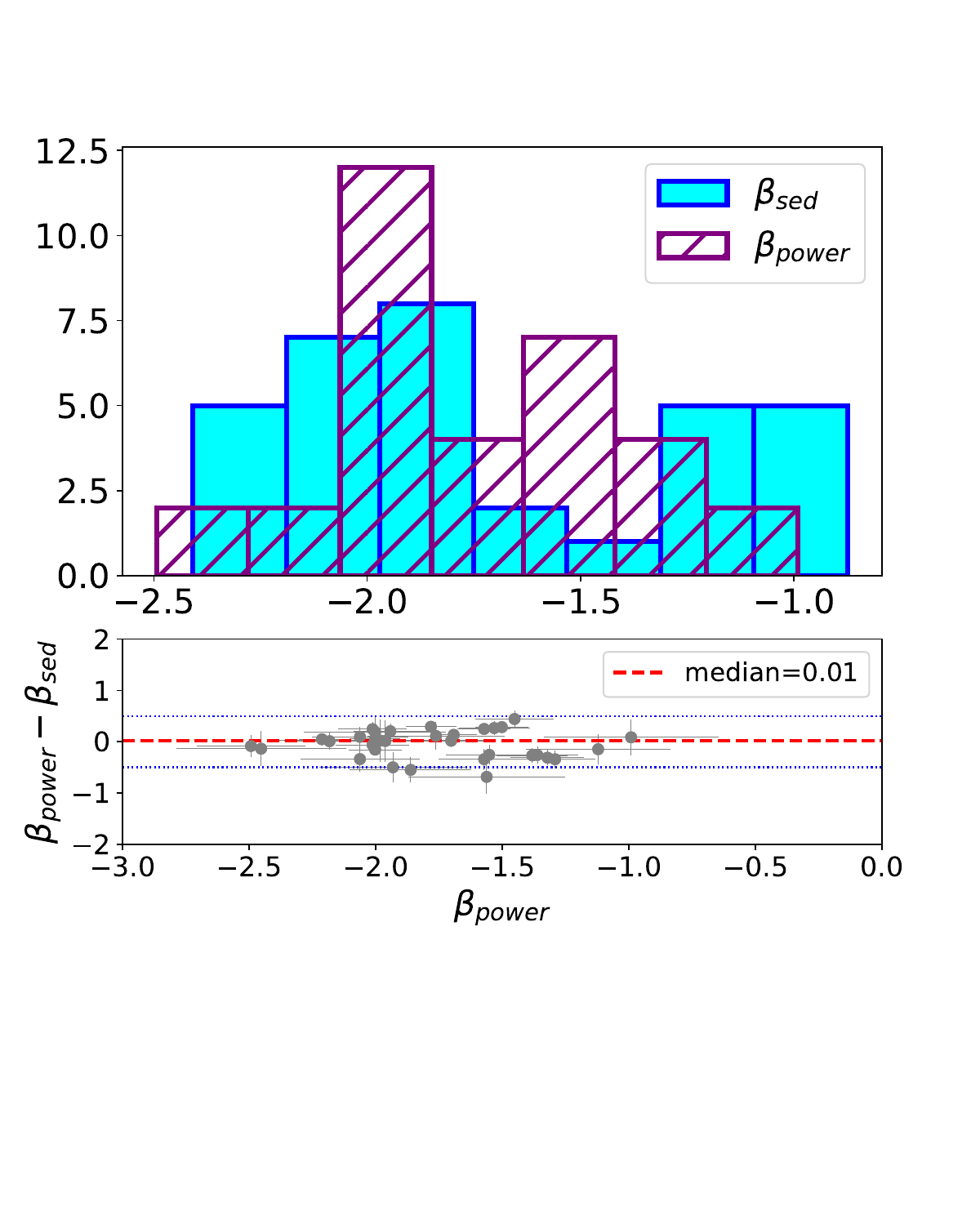}
\caption{Comparing the $\beta_{\mathrm{power}}$ and $\beta_{\mathrm{SED}}$ values. Upper row: The distribution of $\beta_{\mathrm{power}}$  and $\beta_{\mathrm{SED}}$ 
values are shown with the purple and blue histograms, respectively. 
Lower row: The difference between the $\beta_{\mathrm{power}}$ and $\beta_{\mathrm{SED}}$ measurements versus the $\beta_{\mathrm{power}}$ values on the x-axis. The dotted blue lines define the region where the difference between $\beta$ values is below 0.5.}
\label{fig:compare}
\end{figure}

\textbf{As shown in the upper panel of Figure \ref{fig:compare}, the two $\beta$ measurements generally agree, with a median of $|\Delta\beta|=0.2$, comparable to the median combined uncertainty of the two measurements, $\mathrm{median}( \sqrt{\Delta \beta_{\mathrm{SED}}^{2} + \Delta \beta_{\mathrm{power}}^{2} })=0.2$.} 

\textbf{The lower panel of Figure \ref{fig:compare} illustrates the difference between the $\beta_{\mathrm{power}}$ and $\beta_{\mathrm{SED}}$ measurements. Among the 33 galaxies in this sample, $83\%$ have $|\Delta\beta|=|\beta_{\mathrm{power}}-\beta_{\mathrm{SED}}|<0.5$ (blue dotted lines in Figure \ref{fig:compare}), with a general uncertainty of 0.2 (see above). This demonstrates again that these two measurements of UV slope are consistent.}

\textbf{The strong agreement between $\beta_{\mathrm{power}}$ and $\beta_{\mathrm{SED}}$ is expected, as the stellar population models in the SED fitting are fit to the same broadband photometry used to derive $\beta_{\mathrm{power}}$ \citep{reddy2015, Morales2025}. For the rest of the analyses in this paper, we use only the $\beta_{\mathrm{power}}$ values, although the results remain the same if we use the $\beta_{\mathrm{SED}}$ measurements.}


\section{Nebular dust attenuation}
\label{sec:nebular_dust}
 As suggested in \citet{cal94}, for a simple case of a uniform layer of dust between the source and observer, the intensity of dust extinction along line-of-sight can be characterized by optical depth, as follows:

\begin{equation} 
\label{equ:line_ext}
\begin{split}
    f(H\alpha) = f^{0}(H\alpha) \times e^{-\tau_{H\alpha}} \\
    f(H\beta) = f^{0}(H\beta) \times e^{-\tau_{H\beta}}   
\end{split}
\end{equation}

Where $f^{0}$ is the intensity of emission line that would be observed in the absence of dust. Also, $\tau_{H\alpha}$ and $\tau_{H\beta}$ are the optical depths at the wavelengths of $H\alpha$ and $H\beta$, respectively. 
The Hydrogen Balmer emission lines originate from ionized gas (i.e., HII regions) in galaxies and can be utilized to estimate the dust extinction towards these ionizing regions. The nebular dust attenuation of galaxies is usually measured using the ratio of the two lowest-order emission lines of the Balmer series, H$\alpha$ and H$\beta$.

\begin{equation} 
\label{equ:balmer_dec}
    \tau_{B} = \tau_{H\beta} - \tau_{H\alpha} = \ln{(\frac{f(H\alpha) /f(H\beta)}{2.86})}
\end{equation}

where $\tau_{B}$ is defined as the Balmer optical depth \citep{cal94}. The intrinsic ratio of 2.86 corresponds to a temperature $T=10^{4}$ K and an electron density of $n=10^{2}$ cm$^{-3}$ for Case B recombination \citep{Osterbrock_Ferland2006}. $\tau_{B}$ provides a measure of the optical depth of the dust, which, as any optical depth, is independent of the assumed attenuation curve.

\begin{figure}[ht]
\includegraphics[width=\columnwidth]{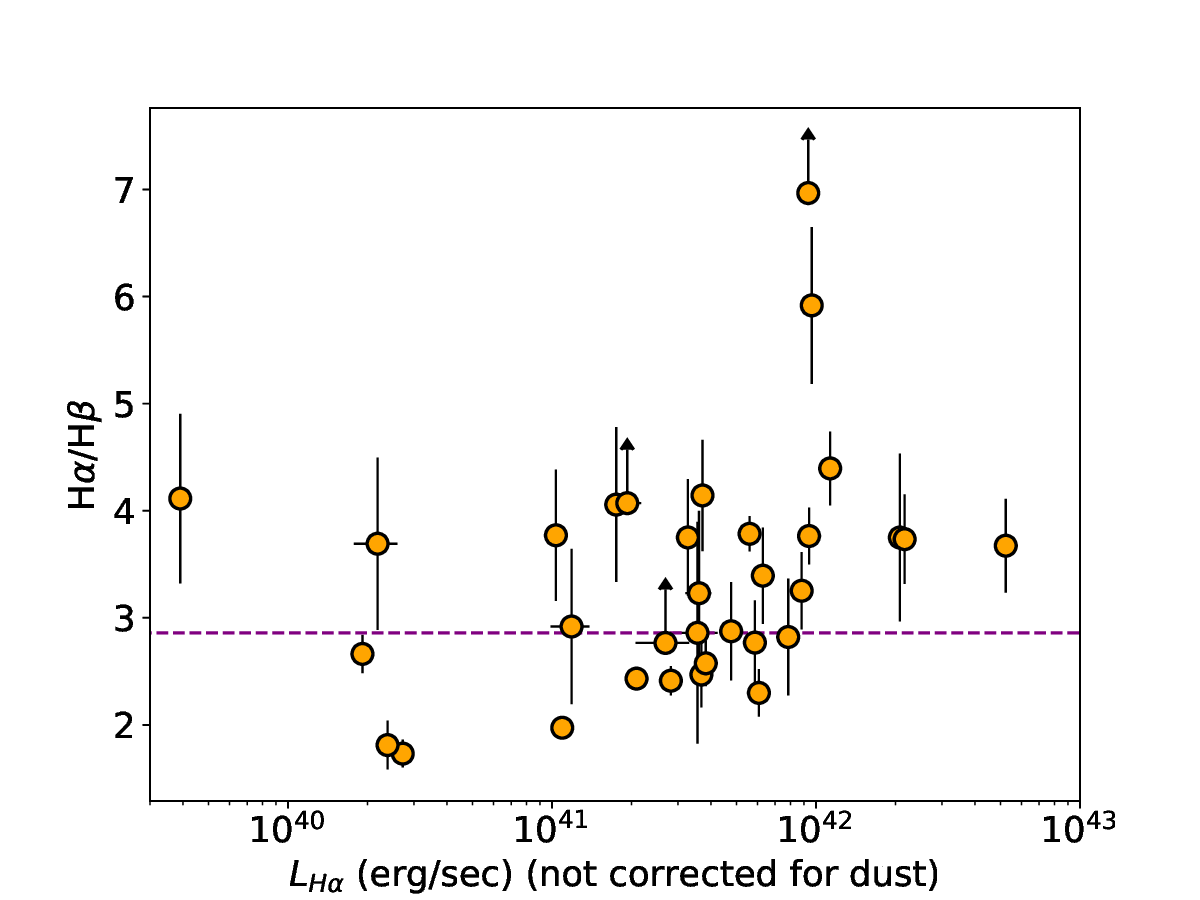}
\caption{Ratio of H$\alpha$ to H$\beta$ emission lines versus lensing-corrected H$\alpha$ luminosity for our sample. 
The emission line fluxes have been adjusted for the underlying Balmer absorption. Additionally, the dashed purple line denotes the theoretical value of 2.86 for H$\alpha$/H$\beta$ under normal ISM conditions without dust. As seen here, a fraction of sources in our sample have the Balmer ratio lower than theoretical value. The arrows represent $3\sigma$ lower limit for the line ratio whenever H$\beta$ is undetected.}
\label{fig:Ha_to_Hb}
\end{figure}

We utilize H$\alpha$ and H$\beta$ fluxes that have been adjusted for Balmer absorption (see Section \ref{line_flux}) to determine the $\tau_{B}$ values for our sample. In some cases, the calculated $\tau_{B}$ turns out to be negative due to the Balmer line ratio falling below the theoretical value of 2.86. We provide a detailed discussion of these negative values in Section \ref{subsec:neb_dust_mass}. Figure \ref{fig:Ha_to_Hb} presents the corrected ratio of H$\alpha$/H$\beta$ for our sample plotted against the H$\alpha$ luminosities. The H$\alpha$ luminosities have been corrected for the lensing magnification, although they remain uncorrected for dust effects. We note that the error bars here are only \textbf{from measurement noise}, and the ratios are subject to systematic uncertainties in the flux calibration differences between the spectroscopic bands where each line is measured and Balmer absorption corrections. For the sources with non-detected H$\beta$, we use $3\sigma$ lower limit (arrows in Figure \ref{fig:Ha_to_Hb}) for their Balmer line ratios.

\section{Stacking}
\label{sec:stack}
In addition to individual measurements, we explore the average behavior of nebular dust with respect to various physical properties by creating composite spectra. These stack spectra are constructed in bins of UV spectral slope and stellar mass. 
Our stacking methodology closely follows the approach outlined in \citet{henry2021}. In summary, each spectrum is first shifted to its rest frame. To mitigate the influence of a few bright sources, we normalize each spectrum with respect to its H$\alpha$ flux. After interpolating all spectra onto a common grid of rest-frame wavelengths, we compute the median of the normalized fluxes at each wavelength. 

\textbf{We also construct the error spectra for the stacks. For each bin of UV spectral slope or stellar mass, we generate 100 random realizations of each individual normalized spectrum within that bin by perturbing the flux according to its observational uncertainties. We then create 100 median stacked spectrum from these random realizations, following the same procedure described above. Finally, the standard deviation of these random stacks in each bin is taken as the error spectrum for the stack spectrum of that bin. For each bin, the H$\alpha$ and H$\beta$ emission lines and their uncertainties are determined by a single Gaussian fit to the median stack and its error spectrum. We note that some studies, such as \citet{henry2021} and \citet{freeman2019}, found that a double Gaussian fit better represents the line profile in their stacked spectra. We also try the double Gaussian fit and we still get the same results within the error bars. }


\begin{figure*}[ht]
\includegraphics[width=\textwidth]{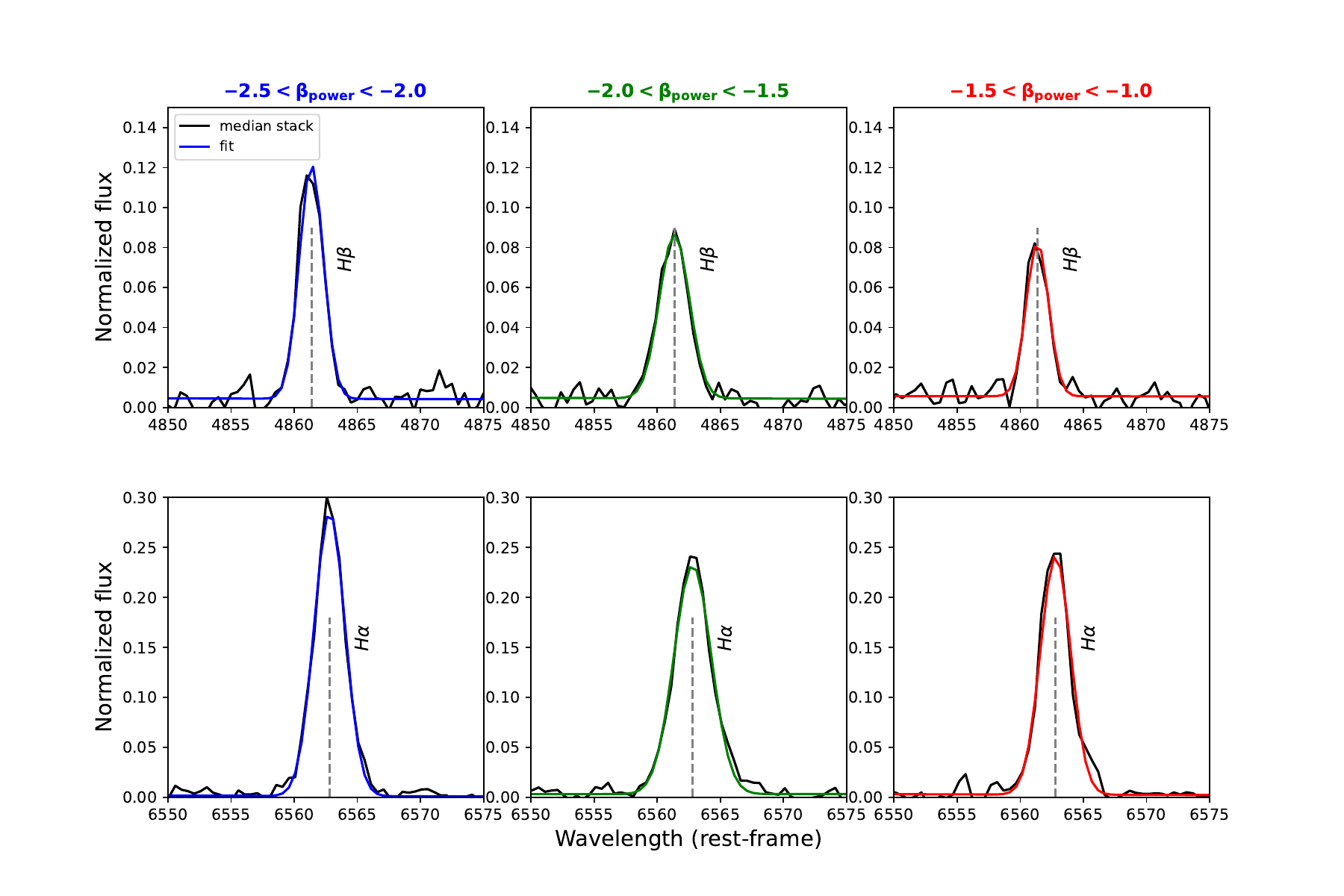}
\caption{Stacked spectra at rest frame. These composite spectra are generated using the median stacking technique (see Section \ref{sec:stack}). Prior to stacking, all individual spectra were normalized to their H$\alpha$ flux. The top (bottom) row exhibits the composite spectra at the H$\beta$ (H$\alpha$) wavelength for three bins of $\beta_{power}$, with the bluest on the left and the reddest on the right. Within each panel, the stack is represented in black, while the best fit is depicted with blue, green, and red lines, corresponding to the blueness of the UV spectral slope bin. The wavelength of the fitted line (H$\alpha$ or H$\beta$) is indicated with a dashed line.}
\label{fig:stack_beta}
\end{figure*}

\textbf{Figure \ref{fig:stack_beta} illustrates the composite spectra organized into bins of $\beta_{power}$. The composite spectra are displayed solely around the wavelengths of H$\alpha$ and H$\beta$ emission lines. 
The three panels are arranged from the bluest $\beta$ bin on the left to the reddest $\beta$ bin on the right. Each panel also shows the best single Gaussian fit, depicted in blue, green, and red, corresponding to the blue-to-red progression of $\beta_{power}$.}

\begin{figure}
\includegraphics[width=\columnwidth]{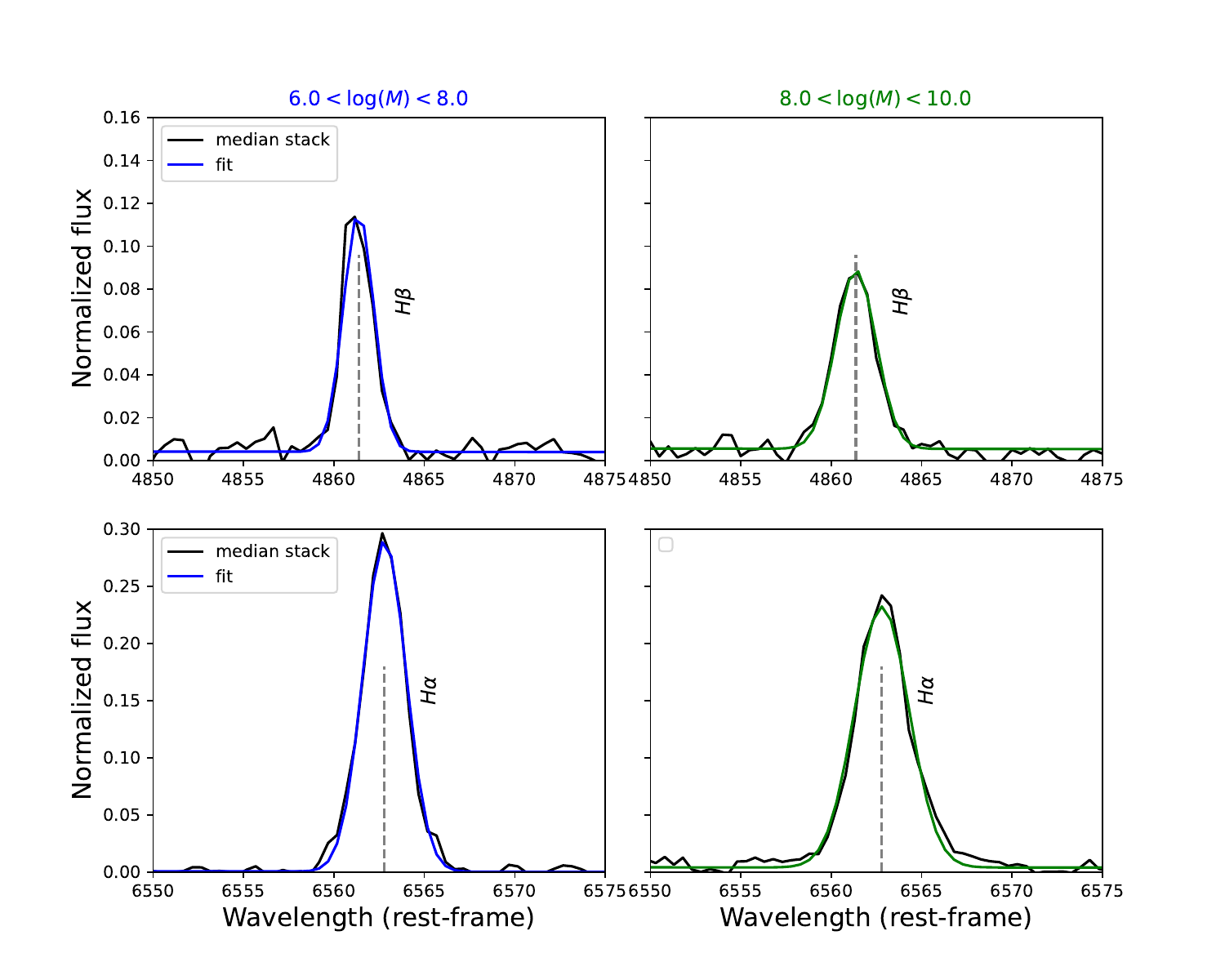}
\caption{The same as figure \ref{fig:stack_beta} for bins in stellar mass, with the lowest mass bin on the left column to the highest mass bin on the right column.}
\label{fig:stack_mass}
\end{figure}
 



\begin{figure*}[ht]
\includegraphics[width=\textwidth]{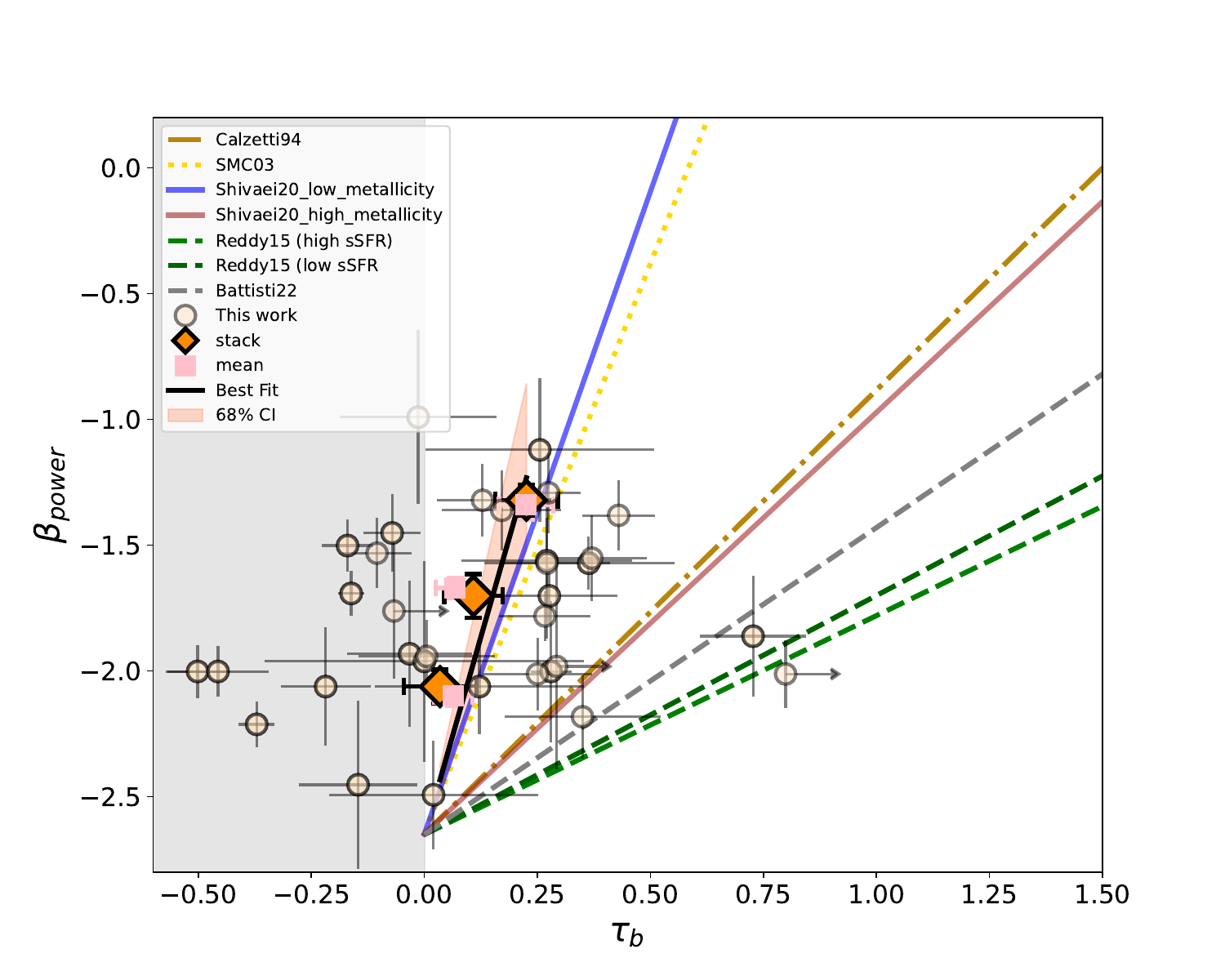}
\caption{This plot illustrates UV spectral slope $\beta$ versus Balmer optical depth $\tau_{B}$. The $\beta$ slopes measured from power-law fitting ($\beta_{power}$) are shown on the y axis. 
 The orange diamonds and pink squares show median stack and average values (see text), respectively. All lines represent different attenuation curves from literature as listed in the legend box. The blue and red solid lines are from \citet{Shivaei2020} for low-metallicty and high-metallicity sources at cosmic noon, respectively. The dotted golden line is for SMC and dashed dotted brown line is for the Calzetti curves. The dashed gray line shows the attenuation curve from \citet{battisti2022} found for a sample of emission-line galaxies at $z\sim1.3$. The dashed green lines are from \citet{reddy2015} for normal star-forming galaxies with low and high sSFR at $z=1.4-2.6$. Our sample is best represented by the SMC curve as well as \citet{Shivaei2020} dust curve from their low-metallicity sample. The black solid line is the best fit to our stacks (orange diamonds). The pink area is the 1-$\sigma$ uncertainty around our best fit.}
\label{fig:beta_balmer_tau}
\end{figure*}

\section{Results}
\label{sec:results}
\subsection{Balmer optical depth and relation with UV spectral slope}
\label{subsec:balmer_depth_UV}

UV photons undergo significant absorption and scattering by dust grains, influencing the overall shape (i.e., the UV spectral slope) and intensity of the UV spectrum observed from a galaxy. Typically, a steeper $\beta$ slope (more negative value) suggests a lower degree of dust attenuation. Consequently, if the $\beta$ slope is sensitive to dust, it is reasonable to expect a correlation with nebular dust attenuation, which is parameterized by $\tau_{B}$. 

Figure \ref{fig:beta_balmer_tau} illustrates the correlation between the UV spectral slope $\beta$ and the Balmer optical depth $\tau_{B}$ for our targets. \textbf{The $\beta$ slopes derived from power-law (see Section \ref{sec:uv-slope}) are displayed in this plot.}  
The individual sources within our sample are represented by orange circles. To illustrate the average trend between $\tau_{B}$ and $\beta$ for low-mass galaxies, we incorporate median stacks in bins of $\beta$ values, which are denoted by orange diamonds.
Additionally, besides regular median stacking, we compute an average of the measured line fluxes within each bin. Within each bin, we normalize the H$\alpha$ and H$\beta$ line fluxes of individual sources by their H$\alpha$ fluxes. Subsequently, we calculate the weighted mean of the normalized H$\alpha$ and H$\beta$ fluxes. The weight for each flux is determined as the inverse square of its uncertainty. The resulting $\tau_{B}$ value of each bin is derived from the ratio of these mean fluxes and they are shown as pink squares. As seen in this figure, the median stack and mean $\tau_{B}$ values agree well.

\textbf{We compute the Spearman correlation coefficient \citep{spearman1904} to assess the significance of the correlation between $\beta$ and $\tau_B$. We find $\rho_{\text{Spearman}} = 0.3$ and $p$-value = 0.2, indicating a weak to moderate correlation that is not statistically significant. While a moderate positive correlation is observed between $\beta$ and $\tau_B$, the lack of statistical significance must be due to intrinsic scatter in the relationship. Indeed, as shown by the orange diamonds and pink squares, stacking the spectra in bins of $\beta$ helps to smooth out the individual variations caused by intrinsic scatter and reveals the underlying trend more clearly.}

\textbf{This intrinsic scatter leads to a large variation in $\beta$ for a given $\tau_B$, such that $\beta$ can range from $-2.5$ to $-1.0$ in the limit as $\tau_B$ $\rightarrow$ $0$. A similar dispersion in $\beta$ values are seen in other studies such as \citet{reddy2015} and \citet{battisti2022}. $\beta$ slope is sensitive to other properties of the galaxy other than dust attenuation, which could cause some of the scatter seen in this relation. Among these galaxy properties, $\beta$ can notably change with the star formation history (SFH), metallicity, and the geometrical arrangement of dust and stars.}


\begin{itemize}
    \item Metallicity: The galaxies in our sample span a stellar mass range of log$(M_{*}/M_{\odot}) \sim 7-9.6$ (and one galaxy with very low stellar mass of log$(M_{*}/M_{\odot}) \sim 6.1$). Based on the mass-metallicity relation at similar redshifts \citep{Sanders2021}, the range of gas-phase metallicity probed by our sample is approximately $0.1-0.4\ \rm{Z\odot}$, assuming the solar metallicity values from \citet{Asplund2021}. To confirm this further, in \citet{gburek23}, we directly measured the metallicity for a subset of our sample using the electron-temperature-sensitive [OIII] $\lambda 4363$ emission line. Our findings demonstrated that these galaxies indeed exhibit low metallicities, with $12+\log(\rm{O/H})_{\rm{direct}}=7.88^{+0.25}_{-0.22}$ ($0.15^{+0.12}_{-0.06}\ \rm{Z\odot}$). Assuming a \citet{cha03} IMF and constant star formation over 100 Myr, the \citet{bru03} stellar population synthesis models indicate a difference in the UV slope of $\Delta\beta<0.2$ for the range of metallicities covered in our sample. Therefore, the influence of metallicity on the intrinsic scatter in Figure \ref{fig:beta_balmer_tau}  is likely minimal for our sample.

    \item Dust and star distribution: The scatter caused by the geometry of dust distribution is expected to be minimal, as the galaxies in our sample are predominantly small and compact. This characteristic leads to minimal dispersion of stars and ionized gas within these galaxies. Further evidence is provided by the observation of significant variation in $\beta$ slope values, ranging from $\beta_{power}=-2.6$ to $-1.0$) when $\tau \rightarrow 0$. As highlighted in  \citet{cal94}, the spatial arrangement of dust and stars cannot account for the considerable scatter in $\beta$ when $\tau \rightarrow 0$. Under such circumstances, we anticipate $\beta$ to converge towards its intrinsic value, typically ranging between $\beta_{int} = -2.6$ and -2.0, depending on specifics such as the initial mass function (IMF), age,  and metallicity of the stellar population model. 

    \item SFH: The other factor that could be contributing to the scatter in the $\beta-\tau_{B}$ relation for our targets is the stochastic SFHs in these low-mass galaxies. 
    Hydro-dynamical simulations \citep{gov12,Trujillo-Gomez2015, sparre2017}, which implement strong stellar feedback (e.g., supernova and stellar winds), predict frequent dramatic changes in the star formation rate of low-mass galaxies on very short time scales of $<10$ Myr (i.e., time scale similar to the O-type star lifetime). Several studies \citep{dominguez2015,sparre2017} have used these simulations with bursty SFHs to examine the effect of variable short-timescale SFRs on various galaxy properties and observational relationships
    
    \textbf{The measurement of the UV spectral slope is primarily governed by the contribution of O-, B-, and A-type stars. A simple comparison of their spectra shows that O-type stars exhibit much bluer UV continua than A-type stars. Consequently, during a burst of star formation, when the stellar population is dominated by young, massive O- and early B-type stars, the integrated UV SED of the galaxy becomes significantly bluer. As the burst subsides and the most massive stars quickly evolve off the main sequence, the UV output becomes dominated by longer-lived A-type stars, which are much redder in the UV. This leads to a rapid reddening of the UV spectral slope following the burst, even in the absence of changes in dust content. Thus, the UV slope is a sensitive tracer of recent star formation activity on short timescales (tens to hundreds of Myr). A similar effect has been demonstrated in \citet{Wilkins2012}, using semi-analytical galaxy formation models. They showed that after prolonged periods of star formation some fraction of the most massive stars will have evolved off the main-sequence, and thus the resulting UV slope is redder. }

\end{itemize}


\textbf{To assess the $\beta$–$\tau_{B}$ relationship in our sample relative to established dust attenuation curves, we consider the well-known Calzetti \citep{cal94} and SMC \citep{gor03} curves in the local universe, as well as the Reddy15 \citep{reddy2015}, Shivaei \citep{Shivaei2020}, and Battisti \citep{battisti2022} curves at redshifts comparable to our sample. We adopt the direct $\beta$–$\tau_{B}$ linear relation from equation 6 of \citet{battisti2022} for their sample of emission-line galaxies at $z\sim1.3$ (dashed gray line in Figure \ref{fig:beta_balmer_tau}), and the relation from Table 2 of \citet{reddy2015} for their sample of normal star-forming galaxies at cosmic noon (dashed green lines in Figure \ref{fig:beta_balmer_tau}). For the Calzetti attenuation curve, we use equation 4 of \citet{cal94} (brown dotted-dashed line in Figure \ref{fig:beta_balmer_tau}). We adjust the intercepts of these linear relations to match the intrinsic slope of our sample, $\beta_{\text{intrinsic}} = -2.65$. This value corresponds to the intrinsic $\beta$ for our fiducial BC03 simple stellar population model with $0.2\ \text{Z}{\odot}$ metallicity and 100 Myr constant star formation (see Section \ref{sec:sed-fit} for justification of these assumptions). This blue $\beta$ also agrees with the bluest $\beta_{\text{power}}$ measured in our sample, independent of any spectral model assumptions.}

\textbf{For the SMC and Shivaei attenuation curves, the $\beta$–$\tau_{B}$ linear relations are not explicitly provided in the original papers, so we derive them ourselves. First, we compute the nebular reddening, $E(B-V)_{\text{neb}}$, over a range of $\tau_{B}$ values using each dust curve and equation \ref{equ:E_BV_neb}.}
\textbf{We then estimate the stellar dust reddening, $E(B-V)_{\text{star}}$, from these $E(B-V)_{\text{neb}}$ values. Observations of local star-forming galaxies suggest enhanced reddening toward the nebular ionized gas compared to the stellar continuum \citep{fanelli1988,cal94,mas-hesse1999,moustakas2006,koyama2018}.
 An illustration of this relationship is provided by \citet{cal00}, who find that $E(B-V)_{\text{star}}[\text{Calzetti}] = 0.44 \times E(B-V)_{\text{neb}}$. A similar result is reported by \citet{Shivaei2020}, who find an average of $E(B-V)_{\text{star}}[\text{Shivaei}] = 0.47 \times E(B-V)_{\text{neb}}$. For the SMC curve, we adopt $E(B-V)_{\text{star}}[\text{SMC}] = 0.37 \times E(B-V)_{\text{neb}}$, following the results of \citet{reddy2020}. Finally, to quantify the impact of dust on the UV slope, we apply these stellar dust reddening values to our fiducial BC03 SED model (described above), measure the resulting UV slope from the reddened SED, and thereby we derive the $\beta$–$\tau_{B}$ relationship for the SMC (dotted golden line in Figure \ref{fig:beta_balmer_tau}) and Shivaei dust curves (blue and red solid lines in Figure \ref{fig:beta_balmer_tau}). To validate the accuracy of this approach, we tested it with the Calzetti curve and found that our resulting $\beta$–$\tau_{B}$ relation agrees exactly with equation (4) of \citet{cal94}.}


\textbf{As shown in Figure \ref{fig:beta_balmer_tau}, both the stacks (orange diamonds) and the mean values (pink squares) from our sample agree well with the $\beta-\tau_{B}$  relations from the SMC (golden dotted line) and the low-metallicity dust curve from \citet{Shivaei2020} (solid blue line).
We fit a line (black solid line) to our stacks, assuming an intrinsic UV slope of $\beta_{0} = -2.65$ (see above), as
\begin{equation}
\beta = 6.3^{+1.7}_{-1.3} \times \tau_B - 2.65.
\end{equation}
As shown in this figure, our best-fit line and the $1\sigma$ scatter around it (pink shading) agree with SMC dust curve, a low-metallicity and low-mass galaxy in the local universe, as well as low-metallicity galaxies at $z \sim 2$, as suggested by \citet{Shivaei2020}.}


Finally, the low-mass galaxies in our sample deviate from the trends seen in more massive star-forming galaxies at high redshifts, as represented by \citet{battisti2022} at $z\sim1.3$ (gray line) and \citet{reddy2015}  at $z=1.4-2.6$. This is likely due to the metallicity differences between our sample and those in these studies. As discussed in \citet{Shivaei2020}, metallicity is the key factor influencing the shape of dust attenuation curves in galaxies. They found that high-metallicity galaxies ($12+\log(\rm{O/H})>8.5$) at $z = 1.4-2.6$ exhibit a shallow attenuation curve similar to the Calzetti starburst curve, while low-metallicity galaxies ($12+\log(\rm{O/H})=8.2-8.5$) have a steeper slope, identical to the SMC curve. This is in agreement with what we found.

\subsection{How does Nebular dust change with stellar mass?}
\label{subsec:neb_dust_mass}
Dust attenuation has been shown to correlate with several galaxy parameters, including metallicity, $H\alpha$ luminosity, equivalent width of H$\alpha$ emission line, and stellar mass \citep{Hopkins2001,Sobral2012, dominguez2013, battisti2022, shapley2022}. Several studies demonstrated that dust attenuation primarily depends on stellar mass \citep{battisti2022,maheson2024}. In these studies, a number of techniques are used to determine dust attenuation, including UV dust attenuation $A_{1600}$ derived from UV spectral slopes $\beta$ \citep{Mclure2018, shapley2022}, the ratio of far-infrared to UV luminosities known as IRX \citep{meu99,bouwens2016,red18}, and nebular dust attenuation from Balmer decrements \citep{dominguez2013,Theios2019,shapley2022,shapley2023}. However, many of these investigations at high redshifts focus on higher stellar mass galaxies with $\log(M_{*}/M_{\odot}) > 9$.

Here, we examine how nebular dust attenuation varies with stellar mass in low-mass galaxies with $\log(M^{*}/M_{\odot}) \leq 9$. To quantify nebular dust, we employ the Balmer optical line ratios (see section \ref{sec:nebular_dust}) and compute nebular dust reddening $E(B-V)_{\text{neb}}$ as follows:

\begin{equation} 
\label{equ:line_dust}
\begin{split}
    f(H\alpha) = f^{0}(H\alpha) \times 10^{-0.4A_{H\alpha}} \\
    f(H\beta) = f^{0}(H\beta) \times 10^{-0.4A_{H\beta}}   
\end{split}
\end{equation}

where $f$ and $f^{0}$ are similar to equation \ref{equ:line_ext}, and A$_{H\alpha}$ and A$_{H_\beta}$ are the nebular dust attenuation measured at H$\alpha$ and H$\beta$ wavelengths, respectively.

\begin{equation} 
\label{equ:A}
\begin{split}
    A_{H\alpha} = E(B-V)_{neb} \times k_{H\alpha} \\
    A_{H\beta} = E(B-V)_{neb} \times k_{H\beta}    
\end{split}
\end{equation}

where $E(B-V)_{neb}$ stands for nebular dust reddening and $k_{H\alpha}$ and $k_{H\beta}$ are the reddening curve evaluated at H$\alpha$ and H$\beta$ wavelengths, respectively. Combining equations \ref{equ:line_dust} and \ref{equ:A}, we can derive $E(B-V)_{neb}$ as follows:

\begin{equation}
\label{equ:E_BV_neb}
\begin{aligned}
E(B-V)_{\mathrm{neb}} &= \frac{2.5}{(k_{H\beta} - k_{H\alpha})} 
   \log\!\left(\frac{f(H\alpha)/f(H\beta)}{f^{0}(H\alpha)/f^{0}(H\beta)}\right) \\
&= \frac{2.5}{(k_{H\beta} - k_{H\alpha})} 
   \log\!\left(\frac{f(H\alpha)/f(H\beta)}{2.86}\right) \\
&= \frac{2.5}{(k_{H\beta} - k_{H\alpha})} 
   \log(e)\,\tau_{B}
\end{aligned}
\end{equation}

Here we used the intrinsic ratio of 2.86 as explained in Section \ref{sec:nebular_dust}.
\textbf{As we concluded in Section \ref{subsec:balmer_depth_UV} that the SMC attenuation curve best represents our sample, we adopt the SMC curve to estimate the $E(B-V)_{\mathrm{neb}}$ values for our galaxies.} 
In the context of star-forming galaxies within the local universe, the \citet{car89} reddening curve of the Milky Way is commonly employed to interpret nebular reddening. In addition, \citet{reddy2020} provided evidence showing that the \citet{car89} curve is also suitable for correcting nebular emission lines for dust effects in galaxies within the redshift range of $1.4\leq z \leq 2.6$. 
\textbf{We note that the SMC dust curve from \citet{gor03} at the wavelengths of the H$\alpha$ and H$\beta$ emission lines is indistinguishable from the Cardelli curve. Therefore, had we used the Cardelli curve to estimate our $E(B-V)_{\mathrm{neb}}$ values, we would have obtained consistent results.} 

\begin{figure*}[ht]
\includegraphics[width=\textwidth]{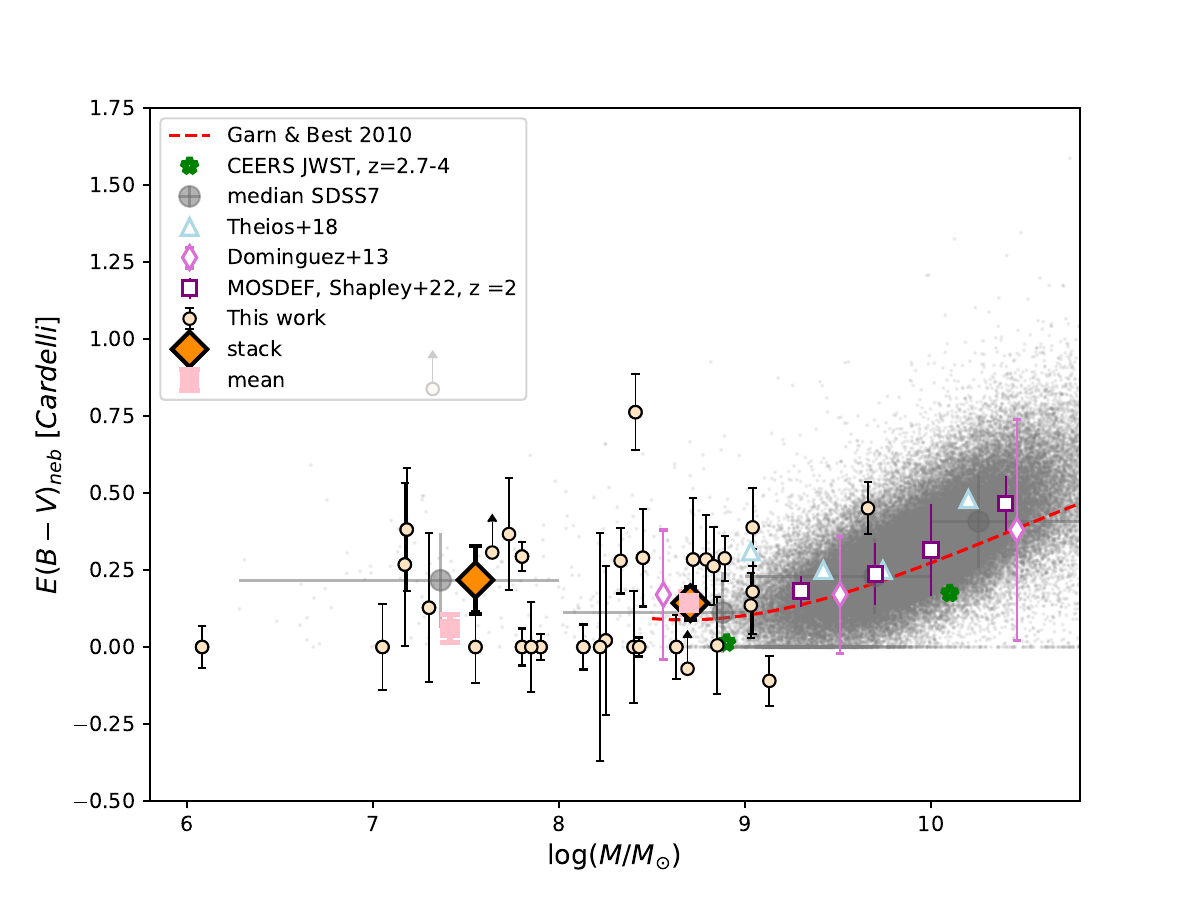}
\caption{The correlation between nebular dust
reddening $E(B-V)_{neb}$ and stellar mass $M_{*}$.
Each individual source in our sample is depicted with orange circles. Similar to Figure \ref{fig:beta_balmer_tau}, the orange diamonds and pink squares represent median stack and average values (see Section \ref{subsec:balmer_depth_UV}), respectively. Gray data points denote local galaxies from the SDSS sample (refer to the text for details ). The red line shows the median of $E(B-V)_{neb}$ values for SDSS galaxies from \citet{garn2010}. The simple median of SDSS values are shown with gray circles.  Additionally, results from the literature at high redshifts are overlaid, including the KBSS sample at $z=2-3$ from \citet{Theios2019} (pink triangle), the WISP sample at $z=0.75-1.5$ from \citet{dominguez2013} (purple diamond), the MOSDEF sample at $z\sim2.3$ from \citet{shapley2022} (purple squares), and the CEERS JWST sample at $z=2.7-4$ from \citet{shapley2023} (green stars). As seen in this plot, our targets align with the flat lower-mass end of the nebular reddening vs. mass relation previously observed for more massive galaxies at similar redshifts. This suggests that the nebular dust versus stellar mass relation shows no significant evolution between local universe and $z=1-3$, extending to low stellar masses of $\log(M_{*}/M_{\odot})=7$.}
\label{fig:mass_E}
\end{figure*}

Figure \ref{fig:mass_E} illustrates the relationship between $E(B-V)_{neb}$ and stellar mass $M_{*}$. Individual galaxies from our sample are displayed with orange circles. The median stack and average reddening in each bin (see section \ref{subsec:balmer_depth_UV}) are represented by orange diamonds and pink squares, respectively. \textbf{The values for our stacks and mean estimates are given in Table \ref{tab:nebular_EBV}}. To provide context and compare with nebular reddening measurements at similar redshifts, we also include results from the literature. The purple squares denote data points from \citet{shapley2022} for a sample of star-forming galaxies from the MOSDEF survey at $z \sim 2.3$. It's worth noting that the galaxies in their study exhibit higher stellar masses ($\log(M_{*}/M{\odot}) > 9$) than those in our sample. The pink triangles, sourced from \citet{Theios2019}, represent data from the KBSS survey, which observes normal star-forming galaxies with $\log(M_{*}/M_{\odot}) > 9$ at $z = 2 - 2.7$. The MOSDEF and KBSS samples consist of similar star-forming galaxies, and as evident in Figure \ref{fig:mass_E}, they display a comparable trend on the $E(B-V)_{\text{neb}}-M_{*}$ plane. The purple diamonds represent data points at lower redshifts ($z=0.75-1.5$) from \citet{dominguez2013}. Additionally, we present the findings from slightly higher redshifts at $z=2.7-4$ from \citet{shapley2023}, drawn from a more recent study based on JWST data from the CEERS collaboration \citep{finkelstein2023}, represented by green stars.

The existing measurements from the literature (mentioned above) collectively indicate that nebular reddening experiences an increase with stellar mass at $\log(M_{*}/M_{\odot}) > 9$. Moreover, the observed trend tends to plateau at lower masses, as one would expect. 
Our targets further validate this trend by extending this relationship down to much lower masses, reaching around $\log(M_{*}/M_{\odot}) \sim 7.0$.

\textbf{We note that a considerable scatter in dust attenuation at the low mass regime can be seen in Figure \ref{fig:mass_E}. Several studies have reported significant scatter in the dust properties of dwarf galaxies, which has been attributed to various galaxy properties, including different star formation histories, dust destruction efficiencies, dust grain size distributions, and chemical compositions. For example, \citet{Remy2014} found a large dispersion in the gas-to-dust ratios of local low-metallicity dwarf galaxies in the Dwarf Galaxy Survey. Similarly, \citet{Hirashita2002} reported a large scatter in the dust-to-gas ratio of blue compact dwarfs at a given metallicity, which they explained through intermittent (i.e., bursty) star formation histories that cause variations in dust destruction efficiency. More recently, simulations by \citet{Choban2024} demonstrated that bursty star formation in low-mass galaxies naturally produces large fluctuations in the dust-to-metal ratio, due to the lack of an equilibrium timescale between dust growth and supernova-driven dust destruction. Taken together, these works suggest that the stochastic nature of star formation in dwarf galaxies plays a critical role in driving the observed scatter in dust content/attenuation and, consequently, dust scaling relations. Relatedly, recent JWST studies have revealed that dwarf galaxies can, in fact, exhibit very large dust attenuation. For example, \citet{Bisigello2023} and \citet{Gandolfi2025} identified a substantial population of dusty star-forming dwarf galaxies up to $z \sim 5$ using NIRCam F200W dropout selection. More than 80\% of their dropout sample consists of low-mass galaxies, with a median stellar mass of $\log(M_\star/M_\odot) \approx 7.3$ at $z<2$, and a median dust attenuation of $A_V \approx 4.9$. They suggested several scenarios to explain this excess dust attenuation, including bursty star formation histories and compact galaxy sizes. Similarly, at low redshift, \citet{Schneider2016} found striking variations in dust content among extremely metal-poor dwarf galaxies. In particular, SBS 0335-052 and I Zw 18 have comparable metallicities, gas masses, and stellar masses, yet differ in their dust-to-stellar mass ratios by a factor of 40–70. They attribute the enhanced dust content of SBS 0335-052 to more efficient grain growth in its cold interstellar medium, facilitated by its compact size and higher gas density. Taken together, these studies demonstrate that dwarf galaxies can show unexpectedly high dust attenuation. Bursty star formation, compact morphologies, and variations in ISM density are all viable physical mechanisms that can drive the large scatter in dust content and attenuation observed in both local and high-redshift dwarf galaxies.}

\textbf{Within the scatter of our $E(B-V)_{\text{neb}}$ values, we also find sources with negative values, which arise from our assumption of Case B recombination. The canonical Case B recombination limit is commonly assumed in the analysis of H II regions. For a typical electron temperature of $T=10^{4}$ K and an electron density of $n=10^{2}$ cm$^{-3}$, a completely dust-free H II region would have $\text{H}\alpha / \text{H}\beta = 2.86$. Although this assumption holds for the vast majority of star-forming galaxies, several studies have shown that the spectra of extreme emission line galaxies (EELGs) can deviate from Case B, with observed $\text{H}\alpha / \text{H}\beta$ ratios falling below the canonical value (i.e, negative $\tau_{B}$). Examples include local low-mass EELGs \citep{Yan17}, Ly$\alpha$ emitters at $z \sim 0.3$ \citet{Atek2009}, and a large sample of $1 < z < 3$ sources from NGDEEP–NIRISS \citep{Pizkal2024}. In addition, many dust attenuation studies of star-forming galaxies at $z > 1$ \citep{reddy2015,Theios2019,Shivaei2020,shapley2022,shapley2023} have seen a non-negligible fraction of sources with negative Balmer decrements. \citet{Scarlata2024} examined an EELG at $z \sim 0.07$ with $\text{H}\alpha / \text{H}\beta$ below the canonical Case B value and argued that Case B may not be valid for such systems. They considered alternatives such as Case A, Case C and resonant scattering of $\text{H}\alpha$ photons in non-spherically symmetric geometries. They also showed that the fraction of sources with negative Balmer decrements correlates strongly with the [O III]/[O II] ($\text{O}_{23}$) ratio. At higher $\text{O}_{23}$ values, a common characteristic of EELGs, the incidence of low $\text{H}\alpha / \text{H}\beta$ ratios increases. For our sample, $\text{O}_{23}$ measurements are available for 73\% (24/33) of the sources, all with $\text{O}_{23} > 1$ up to 35. Among those with $1 < \text{O}_{23} < 5$, roughly 15\% of our sources have negative Balmer decrements that lie well beyond their respective uncertainty ranges, indicating that these values are not attributable to measurement noise alone. This fraction is in great agreement with the predictions of \citet{Scarlata2024} (see their Fig. 7, left panel). For galaxies with higher $\text{O}_{23} >5$, there is a higher fraction of sources with a negative Balmer decrement in our sample ($\sim 40\%$). A detailed investigation of individual sources, as performed by \citet{Scarlata2024}, is beyond the scope of this work. To remain consistent with previous high-redshift dust attenuation studies in how negative Balmer decrements are treated (e.g., Fig. 4 in \citet{shapley2023}; Fig. 1 in \citet{shapley2022}; Fig. 13 in \citet{Theios2019}), we simply set these values at zero (see our Figure \ref{fig:mass_E}).}

\begin{table*}[ht]
\centering
\caption{Comparison of nebular color excess measurements.}
\begin{tabular}{lccc}
\hline\hline
Study & $\log(M_\ast/M_{\odot})$ & $E(B-V)_{\text{nebular}}$ & $E(B-V)_{\text{nebular}}$ error \\
\hline
SDSS (median) & [7.4--8.8] & [0.22, 0.11] & [0.20, 0.10] \\
Our stacks    & [7.6--8.7] & [0.22, 0.14] & [0.11, 0.05] \\
Our mean      & [7.4--8.7] & [0.06, 0.14] & [0.05, 0.04] \\
\hline
\end{tabular}
\label{tab:nebular_EBV}
\end{table*}

To discern the evolution of $E(B-V)_{neb}$ - $M_{*}$ relationships over time, we incorporate data from local star-forming galaxies obtained from the Sloan Digital Sky Survey (SDSS), depicted with gray points.
 For the SDSS galaxies, we use Data Release 7 (DR7) \footnote{https://wwwmpa.mpa-garching.mpg.de/SDSS/DR7/} from \citet{Abazajian2009} and we follow the same selection criteria done by \citet{shapley2022}. In brief, we narrow down our selection to a sub-sample of SDSS galaxies within the redshift range $0.04<z<0.1$ to mitigate the effects of large aperture size. Following \citet{andrews2013}, we 
restrict the SDSS galaxies to those with $5\sigma$ detections in [OII] $\lambda3726$, [OII] $\lambda3729$, H$\beta$, H$\alpha$, and [NeII] $\lambda6584$ as well as a $3\sigma$ detection for [OIII] $\lambda5007$. And finally to exclude AGNs, we adopt the criteria outlined by \citet{kauffmann2003}. \textbf{We calculate the median dust attenuation of SDSS7 galaxies and list them in Table \ref{tab:nebular_EBV}. As seen in Figure \ref{fig:mass_E} and values listed in Table \ref{tab:nebular_EBV}, a significant finding is that the relation between nebular dust attenuation and stellar mass at low masses shows no substantial evolution with redshift, spanning from local SDSS galaxies to $z=1.4-2.6$, as probed by our sample.} This aligns with similar findings reported by \citet{dominguez2013, battisti2022, shapley2023} for more massive star-forming galaxies. Here, we demonstrate that low-mass galaxies at $z=1.4-2.6$ exhibit nebular dust reddening comparable to their counterparts at fixed stellar mass in the local universe. There is a lot of debate in the literature to understand why dust attenuation at a given stellar mass does not evolve with redshift from $z=0$ to cosmic noon. To explain this lack of redshift evolution, \citet{shapley2022} presents a simplified model where dust is evenly distributed, making the optical depth of the dust (and thus the attenuation of the dust) linearly proportional to the surface density of the dust mass. In their model, the absence of evolution in dust attenuation at fixed stellar mass suggests a constant ratio of dust mass surface density to stellar mass. However, given recent findings showing a significant increase in dust mass \citep[i.e., via ALMA observations,][]{Donevski2020,Magnelli2020} and gas content \citep{tacconi2013} with redshift, it is challenging to explain the lack of evolution in dust attenuation.

\section{Summary and Conclusions}
In this paper, we present an analysis of dust attenuation for a sample of 33 lensed, low-mass star-forming galaxies at $1.4<z<2.6$. Exploiting the lensing magnification from three foreground galaxy clusters A1689, M0717 and M114, we explore the dust attenuation at low stellar masses down to $\log(M_{*}/M_{\odot})=7.0$. We use rest-frame optical spectra obtained by Keck/MOSFIRE combined with HST imaging including deep UV data to examine the relationship between dust attenuation, as traced by the Balmer
decrement, and UV spectral slope and stellar mass of galaxies. To this end, we make median stack spectra in bins of stellar mass and UV spectral slope and we  also study how each relation is changing with redshift. Our conclusions are as follow:

\begin{itemize}
\item Nebular dust attenuation measured via the Balmer decrement (i.e., H$\alpha$/H$\beta$) is correlated with the UV spectral slope, albeit with some scatter. Our analysis shows that low-mass star-forming galaxies in our sample at $1.4 < z < 2.6$ seem to follow the $\beta$-$\tau_{B}$ relation as expected from the steep dust attenuation curve of SMC. This $\beta$-$\tau_{B}$ relation seen here applicable for low-mass galaxies is different than what is expected from the widely used Calzetti dust attenuation curve. 
Our $\beta$-$\tau_{B}$ relation is also in agreement with the dust attenuation curve measured for low-metallicity star-forming galaxies at similar redshift from \citet{Shivaei2020}. 
We demonstrate that low-mass galaxies in our sample deviate from the $\beta$-$\tau_{B}$ trends seen for more massive high-redshift star-forming galaxies by \citet{battisti2022} and \citet{reddy2015}.

\item We measure the nebular dust attenuation, $E(B-V)_{neb}$, for our sample, extending the $E(B-V)_{neb}$- stellar mass relation to very low masses of $\log(M_{}/M_{\odot}) \sim 7$. Consistent with expectations, $E(B-V)_{neb}$ decreases with stellar mass, and our results indicate that the relation plateaus at $\log(M_{}/M_{\odot}) < 9$.

\item  We demonstrate that the relationship between nebular dust attenuation, $E(B-V)_{neb}$, and stellar mass in our low-mass galaxies aligns with the relation seen for local SDSS galaxies with similar masses. Although previous studies have reported a lack of redshift evolution in the nebular dust attenuation-stellar mass relation for more massive galaxies ($\log(M{}/M_{\odot}) > 9$) at $1.3 < z < 4$, our study is the first to investigate this relationship at very low stellar masses.
 
Understanding this constant relation between nebular dust attenuation and stellar mass for all star-forming galaxies between the local universe and $z\sim4$ is yet puzzling. A better understanding of how dust content and specially dust surface density is evolving within this redshift range, is needed. A study of resolved dust attenuation with JWST could help to solve this puzzle.

We thank the referee for their valuable comments and suggestions, which have helped improve the paper. \textbf{Some of the data presented in this article were obtained from the Mikulski Archive for Space Telescopes (MAST) at the Space Telescope Science Institute. The specific observations analyzed can be accessed via \dataset[doi:10.17909/ptw6-qt53]{https://doi.org/10.17909/ptw6-qt53}.}

Some of the data presented herein were obtained at Keck Observatory, which is a private 501(c)3 non-profit organization operated as a scientific partnership among the California Institute of Technology, the University of California, and the National Aeronautics and Space Administration. The Observatory was made possible by the generous financial support of the W. M. Keck Foundation. The authors wish to recognize and acknowledge the very significant cultural role and reverence that the summit of Maunakea has always had within the Native Hawaiian community. We are most fortunate to have the opportunity to conduct observations from this mountain.

\end{itemize}

\bibliography{citation.bib}{}
\bibliographystyle{aasjournal}



\end{document}